\documentclass[aps,prb,floatfix,twocolumn,showpacs]{revtex4}
\usepackage{epsfig}
\usepackage{amsmath}
\usepackage{amssymb}
\usepackage{amsfonts}
\usepackage{ dsfont }

\usepackage{color}

\newcommand{\be}{\begin{equation}}
\newcommand{\ee}{\end{equation}}

\renewcommand\Re{\operatorname{\mathfrak{Re}}}
\renewcommand\Im{\operatorname{\mathfrak{Im}}}

\begin{document}


\hsize\textwidth\columnwidth\hsize\csname@twocolumnfalse\endcsname

\bibliographystyle{plain}

\title{Local Spin Susceptibilities of Low-Dimensional Electron Systems
}

\author{Peter Stano$^{1,2}$, Jelena Klinovaja$^1$, Amir Yacoby$^3$, and Daniel Loss$^1$}
\affiliation{$^1$Department of Physics, Klingelbergstrasse 82, University of Basel, Switzerland\\
$^2$Institute of Physics, Slovak Academy of Sciences, 845 11 Bratislava, Slovakia\\
$^3$Department of Physics, Harvard University, Cambridge, Massachusetts 02138, USA}

\vskip1.5truecm
\begin{abstract}
We investigate, assess, and suggest possibilities for a measurement of the local spin susceptibility of a conducting low-dimensional electron system. The basic setup of the experiment we envisage is a source-probe one. Locally induced spin density (e.g. by a magnetized atomic force microscope tip) extends in the medium according to its spin susceptibility. The induced magnetization can be detected as a dipolar magnetic field, for instance, by an ultra-sensitive nitrogen-vacancy center based detector, from which the spatial structure of the spin susceptibility can be deduced. We find that one-dimensional systems, such as semiconducting nanowires or carbon nanotubes, are expected to yield a measurable signal. The signal in a two-dimensional electron gas is weaker, though materials with high enough $g$-factor (such as InGaAs) seem promising for  successful measurements.
\end{abstract}
\maketitle

%

\section{Introduction}

The spin susceptibility quantifies the magnetic polarization arising as a response to a weak magnetic perturbation of the system. In the same way as the density-density response function is a key characteristic of the charge degrees of freedom, the spin susceptibility is fundamental for the description of spin excitations. 

The local spin susceptibility is perhaps most often used in the form of the Rudermann-Kittel-Kasuya-Yosida (RKKY) interaction.\cite{ruderman1954:PR, kasuya1956:PTP, yosida1957:PR} It describes an indirect coupling of localized spins, arising through the spin polarization induced in the band of itinerant electrons. RKKY interaction based effects are too numerous to list---here we only mention issues important from the point of view of semiconductor based\cite{Spin_qubits, Kloeffel_2012} scalable architectures\cite{craig2004:S,rikitake2005:PRB}
for fault tolerant quantum computation.\cite{wang2011:PRA} RKKY has been demonstrated to allow for long distance controlled coupling of a qubit pair,\cite{craig2004:S} to affect the decoherence,\cite{rikitake2005:PRB} and predicted to induce a helical ferromagnetic transition of nuclear spins in a remarkable electron-nuclear feedback mechanism in one dimensions.\cite{braunecker2009:PRL,braunecker2009:PRB} Such a nuclear spin phase transition would substantially 
reduce the dephasing times 
of GaAs spin qubits, while the arising effective helical field has been 
found in helping to establish the Majorana fermion phase\cite{Braunecker_Jap_Klin_2009, alicea2012:RPP} or even induce exotic fractionally charged fermions.\cite{klinovaja2012:PRL, klinovaja2013:CM}

For the effects just mentioned, the spatial structure of the susceptibility is crucial. This structure is known for the model of non-interacting electrons.\cite{vignale}
In low dimensions these results were  extended to include the effects of the spin-orbit coupling,\cite{imamura2004:PRB,chesi2010:PRB} electron-electron interactions in one \cite{egger1996:PRB} and two~\cite{chubukov2003:PRB,chubukov2005:PRB,simon2007:PRL,simon2008:PRB,chesi2009:PRB}
 dimensions, and both together.\cite{zak2012:PRB} However, generally speaking, the interaction effects are very challenging to calculate,\cite{chesi2009:PRB} while often playing decisive role, as e.g. for the above mentioned  helical phase transition.\cite{simon2008:PRB} 

Experimentally, the spin polarization in response to a uniform magnetic field is accessible.\cite{zhu2003:PRL,vakili2004:PRL} However, it corresponds to the limit of the zero wavevector of the static spin susceptibility and does not reveal its spatial dependence. Theoretical calculations predict this structure to be non-trivial, such as having non-standard Fermi liquid features.\cite{zak2012:PRB,governale2012:P} 
Apart from interactions, interesting influence is expected to stem from the spin-orbit interactions. They will induce nonzero transversal components of the spin susceptibility tensor and modulate the diagonal ones in an anisotropic way. This, for example, lifts the restriction on spontaneous ferromagnetic order in low dimensions, imposed by the Mermin-Wagner theorem.\cite{pedrocchi2011:PRL} From a broader point of view, breaking the spin rotational symmetry and spatial isotropy, which the spin-orbit coupling causes, is known to bring up effects potentially useful for spintronics.\cite{tokatly2010:PRB,fabian2007:APS} Elucidation of these effects on the local spin susceptibility calls for an experimental verification, which so far remained out of reach, and which motivates our investigations here. We expect strong impetus for the many-body theory itself once such measurements are realized, the susceptibility serving as a test-bed for different many-body theory approaches.

In this article we assess whether the still missing experimental observation of the  spatial structure of the spin susceptibility could be achieved soon, motivated by the recent advances in the nitrogen-vacancy (NV) center based nanoscale magnetic field detectors.\cite{grinolds2012:U} From this perspective we analyze various materials, geometries and measurement designs, focusing on typical two and one-dimensional semiconductor structures, such as two-dimensional electron gas (2DEG) or semiconducting nanowires. 

Our estimates lead us to the conclusion that with the current NV center detection sensitivity the measurement of the local spin susceptibility is challenging, but possible. The main properties favoring a measurable signal magnitude are low electron density and high $g$-factor of the material. We find one-dimensional systems, such as semiconducting nanowires and carbon nanotubes, to provide a measurable signal even without taking into account interactions, which are expected to further boost the signal by several orders of magnitude. Two-dimensional structures we considered yield much weaker signal. Graphene (single layer, bilayer, pristine or doped) spin susceptibility seems not to be amenable for the measurement. Typical non-interacting 2DEGs fall short of the detection limit by one to three orders of magnitude. We expect the interactions to bring a high $g$-factor 2DEG (InGaAs) to the detection limit, while only the signal refocusing or a very strong exchange based source give hope to enable 
measurement in 
the standard GaAs 2DEG.

The article is organized as follows: In Sec.~II we introduce a formal description and define basic quantities relating to the foreseen spin susceptibility measurement. In Sec.~III we estimate the signal in a source-probe setup specifying to a dipolar (IIIA) and exchange (IIIB) based probe. In Sec.~IIIC we explain the refocusing technique which allows to enhance the signal in a two-dimensional medium and in Sec.~IIID we discuss the effects of interactions. In Sec.~IV we study carbon based nanostructures (graphene and carbon nanotubes). In Sec.~VA we suggest a setup for measuring the short distance structure of the spin susceptibility and in Sec.~VB we consider an alternative to a source-probe measurement and show how the signal can be extracted from the medium equilibrium magnetization noise.  We present our conclusions in Sec.~VI.
  
\section{Spin susceptibility}

\begin{figure}
\includegraphics[width=0.45\textwidth]{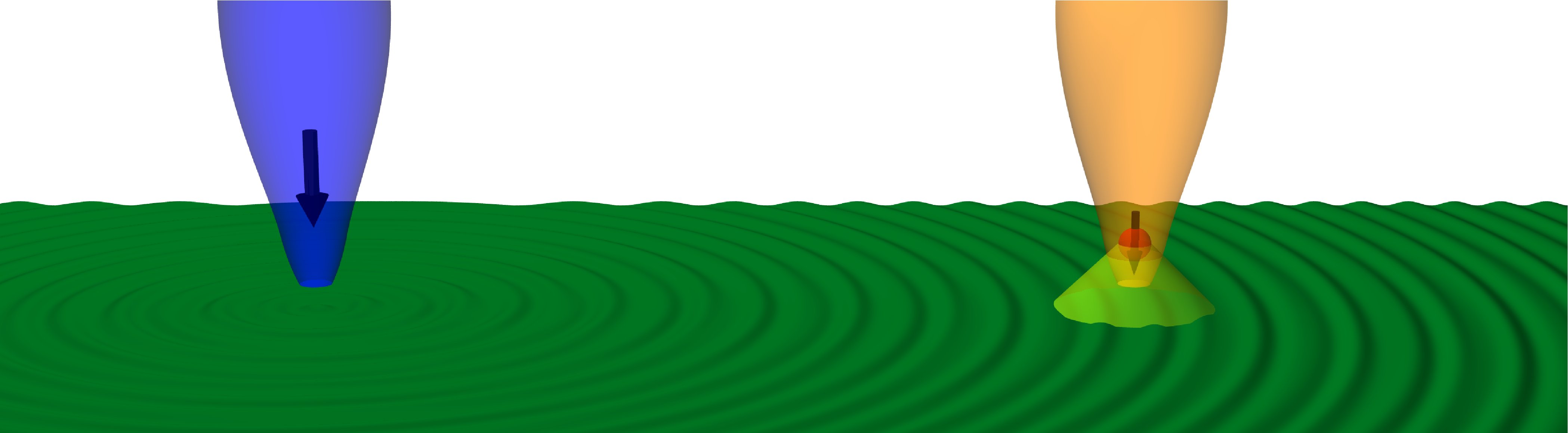}
\caption{(Color online) Source-probe measurement setup. The magnetized source (blue cone with an arrow) excites the medium (green). The spin density excitations (pictured as waves) traversing the medium are detected by the probe containing an NV center (red sphere). The probe collects the signal from the magnetization underneath (light green).}
\label{fig:source-probe-fancy} 
\end{figure}

The spin susceptibility tensor $\chi_{\alpha\alpha^\prime}$ is defined as a linear response quantity relating the induced magnetic moment density ${\bf m}({\bf r},t)$ to an external magnetic field ${\bf B}({\bf r},t)$,
\be
\langle m_\alpha ({\bf r}, t) \rangle = -\mu_\alpha \mu_{\alpha^\prime}\int {\rm d} {\bf r^\prime} {\rm d} t^\prime \chi_{\alpha\alpha^\prime}({\bf r},{\bf r^\prime}; t-t^\prime) B_{\alpha^\prime}({\bf r^\prime},t^\prime).
\label{eq:definition chi general}
\ee
The spatial integration goes over the medium volume, while the time (and frequency, below) integration is over the whole real axis. The same integration limits are assumed in further, unless specified explicitly. 
The Greek indices label Cartesian coordinates, $\mu_\alpha$ is the particle magnetic moment (in general anisotropic), and the angular brackets denote an expectation value $\langle X \rangle = {\rm tr} (\rho X)$ taken with the system density matrix $\rho$.
The relation between a magnetic field produced by the source and a resulting magnetic moment tested by the probe, Eq.~\eqref{eq:definition chi general}, suggests a straightforward way to measure the spin susceptibility in a source-probe experiment. In the following we will analyze mostly such a setup, depicted in Fig.~\ref{fig:source-probe-fancy}. There a magnetized tip is the source of 
a local magnetic field, which excites the medium. 
The source couples to the medium via the Zeeman interaction,
\be
H_I = - \int {\rm d} {\bf r}\,\, {\bf m}({\bf r}) \cdot {\bf B}({\bf r},t).
\label{eq:perturbation dipolar}
\ee
The magnetization of the medium ${\bf m}({\bf r})$  is proportional to the spin polarization $\boldsymbol{\rho}^s({\bf r})$, $m_\alpha ({\bf r})=-\mu_\alpha \rho^s_\alpha ({\bf r})$, where $\boldsymbol{\rho}^s({\bf r}) = \rho({\bf r}) {\bf s}$ with $\rho({\bf r})$ is the particle density operator and $\hbar{\bf s}$ is the spin operator. 
A typical example of a ``medium'' is a two-dimensional electron gas in a semiconductor heterostructure, for which $\mu_\alpha=g \mu_B$, with $g$ the effective $g$-factor, $\mu_B$ the Bohr magneton, ${\bf s}=\boldsymbol{\sigma}/2$, and $\boldsymbol{\sigma}$ the vector of Pauli matrices. 
The emerging spin polarization spreads to large distances, in a form described by the medium spin susceptibility. It is probed locally by an NV based sensor located at another atomic force microscope (AFM) tip. This way, the spatial structure of the spin susceptibility on lengthscales down to the source/probe spatial resolution can be inferred. 

For  perturbations periodic in time 
${\bf B}({\bf r},t) = {\bf B}({\bf r},\omega) \cos(\omega t)$,
it is more convenient to work with the Fourier transform of Eq.~(\ref{eq:definition chi general}),
\be
\langle m_\alpha ({\bf r}, \omega) \rangle = -\mu_\alpha \mu_{\alpha^\prime} \int {\rm d} {\bf r^\prime} \chi_{\alpha\alpha^\prime}({\bf r},{\bf r^\prime}; \omega) B_{\alpha^\prime}({\bf r^\prime},\omega),
\label{eq:chi monochromatic}
\ee
where the spin susceptibility in the frequency space is
\be
\chi_{\alpha\alpha^\prime}({\bf r},{\bf r^\prime};\omega) = \int {\rm d} t\, \chi_{\alpha\alpha^\prime}({\bf r},{\bf r^\prime};t) e^{ {i} \omega t}.
\label{eq:definition FT time}
\ee
It fulfills the Kramers-Kronig relation between the real and imaginary parts,  
\be
\begin{split}
\Re\chi_{\alpha\alpha^\prime}({\bf r},{\bf r^\prime};\omega) = \frac{1}{\pi} P \int {\rm d} \omega^\prime \frac{\Im\chi_{\alpha\alpha^\prime}({\bf r},{\bf r^\prime};\omega^\prime)}{\omega^\prime-\omega}.
\end{split}
\label{eq:Kramers-Kronig}
\ee
Here, $P$ defines the principal value integral.

We will mostly consider the response to a static perturbation, given by Eq.~\eqref{eq:chi monochromatic} with $\omega=0$. We note that the static susceptibility $\chi_{\alpha\alpha^\prime}({\bf r},{\bf r^\prime}; \omega=0)$ is purely real. In addition, from now on we restrict ourselves to a diagonal component of the susceptibility
and assume that the equilibrium state has space translation symmetry. This allow us to introduce a translationally invariant scalar function,
$
\chi({\bf r}-{\bf r^\prime}, t-t^\prime) \equiv
\chi_{\alpha\alpha}({\bf r},{\bf r^\prime} ; t-t^\prime),
$

For a non-interacting system with spin rotational invariance, the spin susceptibility is equal to the density-density response (the Lindhard function). The long-distance static susceptibility of a non-interacting  system of particles with quadratic dispersion at zero temperature is given by\cite{vignale} 
\be
\chi(r) \approx c_d n_d N_d \sin(2k_F r-d \pi/2) (k_F r)^{-d}.
\label{eq:non-interacting result}
\ee
Here $d=1,2$ is the dimension of the system, and we introduced $\chi(r)\equiv\chi({\bf r},\omega=0)$, a notation we use also below. The geometry parameters $c_d$ are defined as $c_1=\pi/4$ and $c_2=1$. The Fermi wavevector $k_F$ corresponds to the Fermi energy $\epsilon_F=\hbar^2 k_F^2/2m$, with $m$ the effective mass. The electron density per spin $N_d$ and the density of states at the Fermi energy $n_d$, are related by $n_d=\partial N_d/\partial \epsilon_F=dN_d/2\epsilon_F$, and read
\be
n_1=m/\pi\hbar^2 k_F, \qquad n_2=m/2\pi \hbar^2.
\label{eq:density of states}
\ee
The two previous equations combined lead to the following expression for the susceptibility,\cite{vignale}  
\be
\chi(r) \approx \frac{m}{4d\pi^d\hbar^2}k_F^{d-2} r^{-d} \sin[2k_F r -d\pi/2].
\label{eq:non-interacting result2}
\ee
We will use Eq.~\eqref{eq:non-interacting result2} further on for scaling estimates. Therefore, the results for the expected signal magnitude we plot are to be taken as a limit from below, since, as mentioned in the Introduction, the interaction effects are expected to boost the spin susceptibility (see Sec.~\ref{sec:interactions} for the enhancement magnitude discussion).

\section{Source-probe setup}

\begin{figure}
\includegraphics[width=0.45\textwidth]{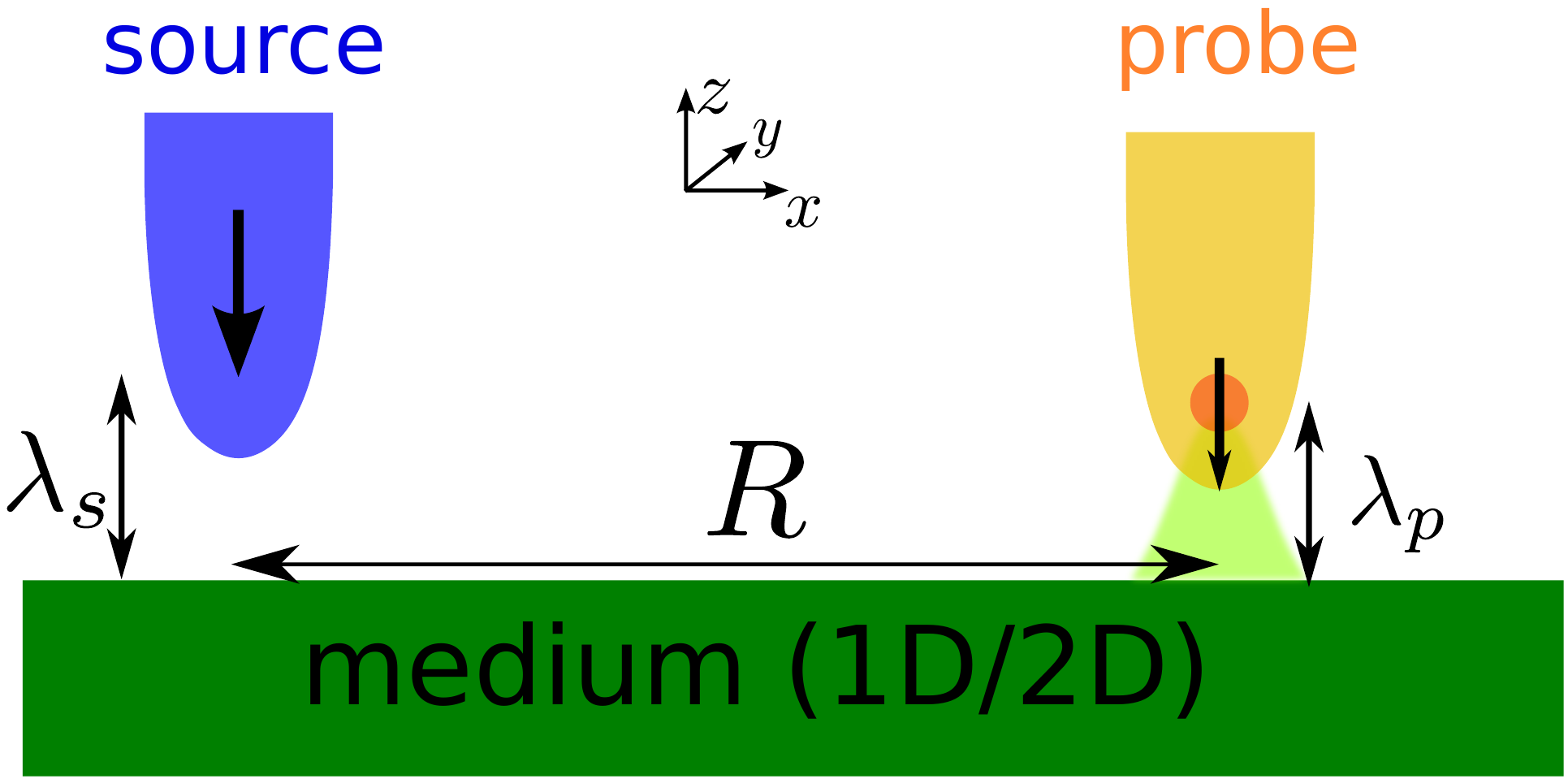}
\caption{(Color online) The source-probe setup redrawn from Fig.~\ref{fig:source-probe-fancy} showing the coordinate system and distance parameters: $\lambda_s$, $\lambda_p$, and $R$ are, respectively, the source-medium, probe-medium, and source-probe in-medium distances. The local magnetic field is produced by the source (blue) placed on the distance $\lambda_s$ from the surface. The magnetic field is measured by the probe (yellow) placed  on the distance $\lambda_p$ from the surface. The probe measures the local magnetic field proportional to the spin susceptibility $\chi(R)$, where $R$ is the distance between the probe and the source.}
\label{fig:source-probe} 
\end{figure}

We now analyze in detail the source-probe setup. The coordinate system and parameters are shown in Fig.~\ref{fig:source-probe}. The dipole field of a magnetic tip serves as the source. In the medium this field is localized over linear distance $\lambda_s$ being the sum of the tip width and its distance to the medium. The induced spin polarization spreads
in the medium according to its spin susceptibility. At distance $R$ (and, possibly, with a controllable time delay for a time dependent source) the probe detects the local spin accumulation by collecting the dipolar field of the magnetic moment induced in an area with linear dimension $\lambda_p$, being the distance of the probe from the medium. This way, the spin susceptibility can be, in principle, mapped out in both space and time variables.

The measurement is, however, by no means straightforward. First of all, if the spatial structure of the susceptibility is aimed at (rather than a response to a uniform field, which has been so far the only experimentally available characteristic of the susceptibility in most of the cases), both source and the probe resolutions have to be below the susceptibility  natural lengthscale. The latter is, as follows from Eq.~\eqref{eq:non-interacting result}, set by the Fermi wavelength, typically tens of nanometers in a semiconductor. Second, though such small magnetic sources are available, scaling the probe down necessarily makes the signal weaker. Third, the dipolar field of the source magnet adds to the field originated in the medium and it must be assured the former is negligible compared to the latter, or the two need to be discriminated, by some scheme identifying a weak signal in a large background. Finally, to learn useful information about the time/frequency structure of the susceptibility, the detection 
time 
resolution must be below the inverse of the natural frequency scale of the susceptibility, typically 1 ps (corresponding to a 1 meV bandwidth/Fermi energy). We will in further assess the susceptibility measurements for different materials and geometries, with the above list of possible issues in mind. 

\subsection{Dipolar field source}

To estimate the signal seen by the probe, we assume the source to be a magnetic moment ${\bf M}$. Its dipolar magnetic field at distance  ${\bf r}$ is defined by tensor ${\bf T}$,
\be
{\bf B}({\bf r})=\frac{\mu_0}{4 \pi} \left( 3 \frac{{\bf r}\cdot {\bf M}}{r^5}{\bf r} - \frac{\bf M}{r^3} \right) \equiv {\bf T}({\bf r})  {\bf M},
\ee
with the permeability $\mu_0=4\pi \times 10^{-7}$ kg m C$^{-2}$. We adopt a coordinate system with the origin in the medium such that the source is at ${\bf l_s}=(0,0,\lambda_s)$ and the probe at ${\bf l_p}=(R,0,\lambda_p)$. The total field at the probe is a sum of two contributions, ${\bf B}_{\rm tot}={\bf B}_0+{\bf B}_t$: the background field ${\bf B}_0$, which is the source dipolar field 
\be
{\bf B}_0 = {\bf T}({\bf l_p}-{\bf l_s})  {\bf M}, 
\ee
and the signal field ${\bf B}_t$ which is due to the spin accumulation transported in the medium,
\be
{\bf B}_t = -\int {\rm d} {\bf r}\, {\rm d} {\bf r^\prime}\, {\bf T}({\bf l_p}-{\bf r})  \overline{\boldsymbol{\chi}}^\prime({\bf r}-{\bf r^\prime})  {\bf T}({\bf r^\prime}-{\bf l_s})  {\bf M}.
\label{eq:signal general}
\ee
Here we introduced a tensor $\overline{\boldsymbol{\chi}}^\prime_{\alpha\alpha^\prime}=\mu_\alpha \mu_{\alpha^\prime} \chi_{\alpha\alpha^\prime}$. We note that if the source magnet can not be considered point-like, ${\bf M}$ should be replaced by a magnetization density ${\bf m}({\bf r^{\prime\prime}})$ and the right hand side of Eq.~\eqref{eq:signal general} should be integrated also over ${\bf r^{\prime\prime}}$ spanning the magnet volume. 

To characterize a measurement feasibility we introduce two figures of merit.
One is the signal absolute magnitude, $B_t$. It is to be compared with the demonstrated NV center detection limit in the order of tens of nanoTesla at ambient conditions. Second is the signal to background ratio, $\gamma=B_t/B_0$. Because a static magnetic field background is irrelevant for the NV sensor, we introduce also a modified coefficient $\gamma^\prime$,
\be
\gamma^\prime = \partial_{\lambda_s} B_t / \partial_{\lambda_s} B_0,
\label{eq:SNRp1}
\ee
which we define as the ratio of changes of the signal and the background upon changing the source to medium distance. Since it is assumed that  $R\ll\lambda_s$ holds in the measurement setup, the signal to background ratio is much higher for the change (rather than the absolute value) of the magnetic field with respect to the variation of the source to medium distance, offering a lock-in technique. 

An alternative lock-in technique is based on time modulations of the electron density, e.g. by a back gate. Such gating does not change the background field ${\bf B}_0$, 
allowing to separate the signal from background once the former is measurable, no matter how big is the latter. We note that changes in the Fermi wavevector as small as $\delta k_F = \pi / 2 R$ revert the spin susceptibility, and thus the signal, sign, see Eq.~\eqref{eq:non-interacting result2}.
This lock-in technique is optimal from our point of view, as it makes the background field value irrelevant. Nevertheless,
we will give the coefficients $\gamma$ and $\gamma^\prime$ to illustrate the 
ratio of the signal and background contributions to the field at the probe.

We now make several approximations to evaluate Eq.~\eqref{eq:signal general} for a static response. We assume the distance between the source and probe $R$ is the largest scale in the system, $R\gg \lambda_p, \lambda_s,1/k_F$. It allows us to use the long distance limit for the susceptibility. We neglect the in-plane components of the source field, as being much weaker than the z-component. Neglecting, in addition, the spin-orbit effects, the resulting spin accumulation is along ${\bf \hat{z}}$ everywhere. We expect the error stemming from these approximations to be exceeded in importance by our uncertainty about the microscopic shape of the source and its magnetic field. Because of this lack of information, the results that follow are expected to contain unknown tip-geometry dependent prefactors of the order of one, which we indicate by using the $\sim$ sign. With these simplifications we get the signal magnitude as
\be
B_t \sim - \frac{\Omega_d^2}{16\pi^2} (g\mu_B \mu_0)^2 M (\lambda_s \lambda_p)^{d-3} \chi(R) \Lambda_d,
\label{eq:Bp1}
\ee
with constants $\Omega_1=2$, $\Omega_2=\pi$, so that $A_d(x) = \Omega_d x^d$ is a volume of a $d$-dimensional sphere of radius $x$. Further, 
\be
\gamma \sim \frac{\Omega_d^2}{4\pi} (g\mu_B)^2 \mu_0 (\lambda_s \lambda_p)^{d-3} R^3 \chi(R)\Lambda_d.
\label{eq:SNR1}
\ee
and finally $\gamma' \approx (1-d/3) (R/\lambda_s)^2 \gamma$. These results can be understood as follows. The probe sees a dipole field $B_t \sim \Lambda_d M_p T_{zz}(\lambda_p \hat{\bf z})$ of an induced magnetic moment $M_p \sim m_p A_d(\lambda_p)$. The dimensionless geometric factor $\Lambda_d$ is discussed below. The magnetization density is proportional to the total ``flux'' of the source field in the medium and the susceptibility, $m_p \sim - \Phi_s (g\mu_B)^2 \chi(R)$. The flux is an area integral of the source field, $\Phi_s \sim A_d(\lambda_s) B_s(0)$. Finally, the field under the source is the dipole field $B_s(0) \sim T_{zz}(\lambda_s \hat{\bf z}) M$. 

In going from Eq.~\eqref{eq:signal general} to Eq.~\eqref{eq:Bp1} we used dimensional analysis to single out the natural parameters dependence. Doing so, the signal is parameterized by a dimensionless factor $\Lambda_d$, which is a function of $\lambda_p k_F$. This is consistent with the limit of large $R$ and the leading corrections to the results below are linear in the small parameters $\lambda_p/R$ and $1/k_F R$. Next, we specify $\Lambda_d$.

The goal is to maximize both the signal and the signal to background ratio.
As the latter is independent of $M$, it is beneficial to use a larger $M$ and bring it as close to the medium as possible. The description of the magnet as a point dipole holds only at distances larger than its linear dimension. In another words, the achievable source magnetic field is limited by the magnet remanent field $B_r=(\mu_0/4\pi) m_r$, with $m_r$ the material magnetization density (typical remanent fields of hard ferromagnets are $B_r = 0.3-1.5$ T). We therefore put $M \sim \lambda_s^3 m_r$. The size of the magnet which produces the maximal signal is then set by the spin susceptibility wavelength $\lambda_s^{\rm opt}=\pi /2 k_F$. This optimal design means the source magnet is designed to produce a maximal possible field, $B_r$, within only a flux 
tube with a diameter $\lambda_s^{\rm opt}$. This is approximately achievable using a prolonged magnetized pillar with the tip width of the order of $\lambda_s^{\rm opt}$. Deviations from the optimal design of the source suppress the signal:  if the source is made smaller the signal diminishes trivially; if the source is made larger, the signal diminishes because of the susceptibility sign oscillations. We do not articulate this suppression any further, as it depends on the tip geometry details.

With the optimal source described above, we can assume the magnetization under the probe is of the form $m_p({\bf r}) \approx m \cos[2 k_F (x-R)]$, which allows us to calculate the dimensionless factor $\Lambda_d$ in the limit of large $R$. We get $\Lambda_2(\alpha) = 2\alpha \exp(-\alpha)$, with $\alpha=2k_F\lambda_p$, whereas we state only the limits in one dimension (see App.~\ref{app:O} for details), $\Lambda_1(0)=1$, $\Lambda_1(\alpha\gg1) \approx \sqrt{\pi \alpha^3/2} \exp(-\alpha)$ If the probe is far away from the medium, the signal is collected from a large area. As a result, it is exponentially suppressed, due to  the  sign oscillations of the magnetization. In one dimensions $\Lambda_1$ grows monotonically upon diminishing the distance between the probe and the medium. In two-dimensional systems, $\Lambda_2$ goes through a maximum at $\lambda_p=1/2k_F$, and decays at small distances, $\Lambda_2(0)=0$. The main reason is that a field of a planar magnet saturates close to the medium,\cite{note_2} 
whereas that of a line magnet diverges---within our model---the one (two) dimensional description of the wire (2DEG) is valid only at distances from it that are larger than the transversal dimension $w$ of the structure, so that the presented formulas are limited to $\lambda_p, \lambda_s \gtrsim w$. 

 Taking together all factors discussed above, we rewrite  Eq.~\eqref{eq:Bp1} as
\be
B_t \sim B_r \frac{\Omega_d^2}{2^{4+d} \pi d}  \mu_0 (g \mu_B)^2  \frac{m}{\hbar^2 k_F^2} \lambda_p^{d-3} \frac{1}{R^{d}} \Lambda_d(2 k_F \lambda_p),
\label{eq:Bp2}
\ee
where we suppressed the sine-like oscillating factor from the susceptibility.\cite{note_3} Similarly we get 
\be
\gamma \sim -\frac{\Omega_d^2}{2^{1+d}\pi^4 d} \mu_0 (g \mu_B)^2   \frac{m}{\hbar^2} \frac{R^{3-d}}{\lambda_p^{3-d}} k_F \Lambda_d(2 k_F \lambda_p). 
\label{eq:SNR2}
\ee
 These formulas allow us to estimate how the figures of merit depend on material and setup parameters: The stronger the source magnet and the material $g$-factor, and the closer the probe to the medium, the better.
On the other hand, the scaling on the Fermi wavevector and the source-probe distance is opposite for the two figures of merit. 

\begin{figure}
\includegraphics[width=0.45\textwidth]{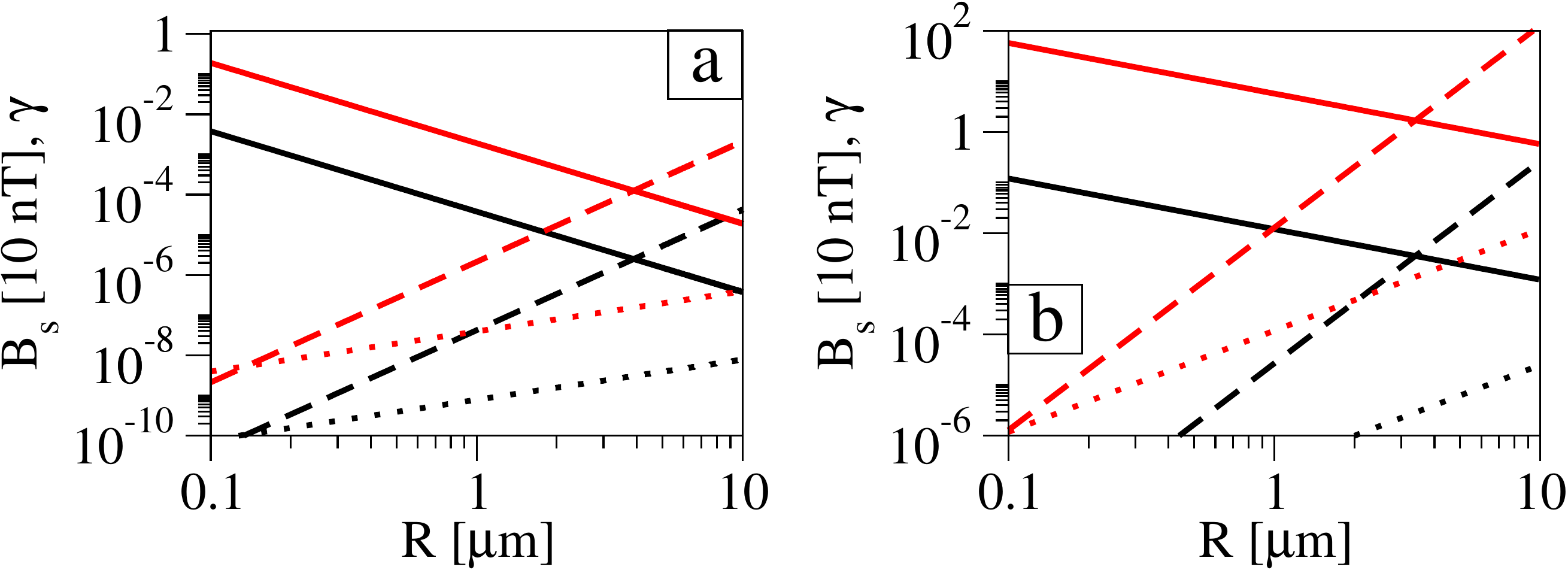}
\caption{(Color online) Figures of merit for a dipolar source in a) 2DEG and b) nanowire. Plotted are: the signal divided by 10 nT (solid lines), and signal to noise ratios, $\gamma$ (dotted), and $\gamma^\prime$ (dashed) for GaAs (black) and InGaAs (red). We used $B_r=1$ T, $1/k_F=50$ nm and $\lambda_p=10$ nm for all materials and geometries. We also used $g=-0.44$, $m=0.067 m_e$, and $E_F=0.23$ meV for GaAs, and $g=-3.9$ (2D) and $g=-12$ (1D), $m=0.043 m_e$, and $E_F=0.35$ meV for InGaAs.
}
\label{fig:dipole-compar}
\end{figure}

In Fig.~\ref{fig:dipole-compar} we illustrate the measurement feasibility for two typical III-V semiconductors. Results given there show that at a distance of the order of a micron from the source the signal falls short by 2-4 orders of magnitude for 2DEG samples. On the other hand, a one-dimensional wire with a relatively high $g$-factor seems very promising. We note, however, that the interactions are expected to generally increase the susceptibility, which might considerably improve the signal measurability (we demonstrate this quantitatively for one-dimensional samples below.) In addition, the signal in two-dimensional samples might be enhanced by refocusing (see below). 

Interestingly, assuming all parameters fixed, the ratio of the figures of merit for a one- and two-dimensional medium boils down to (neglecting numerical prefactors and functions $\Lambda_d$)
\be
\frac{B_t^{\rm 2D}}{B_t^{\rm 1D}}, \frac{\gamma^{\rm 2D}}{\gamma^{\rm 1D}} \sim \frac{\lambda_p}{R}.
\label{eq:non-int-ratio}
\ee
Therefore, a one-dimensional medium is generally preferred if $R\gg\lambda_p$. The dependence of $\lambda_p$ appears because of the area from which the signal is collected,  scaling as $\lambda_p^d$, while $R$ appears because the susceptibility is at large distances inversely proportional to $R^d$.
Apart from this area scaling, a one-dimensional wire is directly accessible for both the source and the probe, while a two-dimensional electron gas is buried tens of nanometers below the material surface, limiting both $\lambda_p$ and $\lambda_s$ from below. This practical issue makes a substantial difference as the dipolar fields drop quickly with these distances.

\subsection{Contact source}

In this section we consider an alternative source, based on a local exchange, rather than dipolar magnetic interaction. A magnetic atom fixed at an AFM 
tip at position ${\bf l_s}$ interacts with the medium through the Hamiltonian
\be
H_I =  \beta \, {\bf I} \cdot \boldsymbol{\rho}^s({\bf l_s}).
\label{eq:perturbation exchange}
\ee
Here $\hbar{\bf I}$ is an atom spin operator, and $\beta$ parameterizes the $s(p)$-$d$ exchange interaction strength. Typical values for $\beta$ for a Mn impurity in a zinc-blende structure semiconductor are 9 meV nm$^3$ for electrons and $-15$ meV nm$^3$ for holes.\cite{furdyna1988:JAP,dietl2001:PRB}

\begin{figure}
\includegraphics[width=0.45\textwidth]{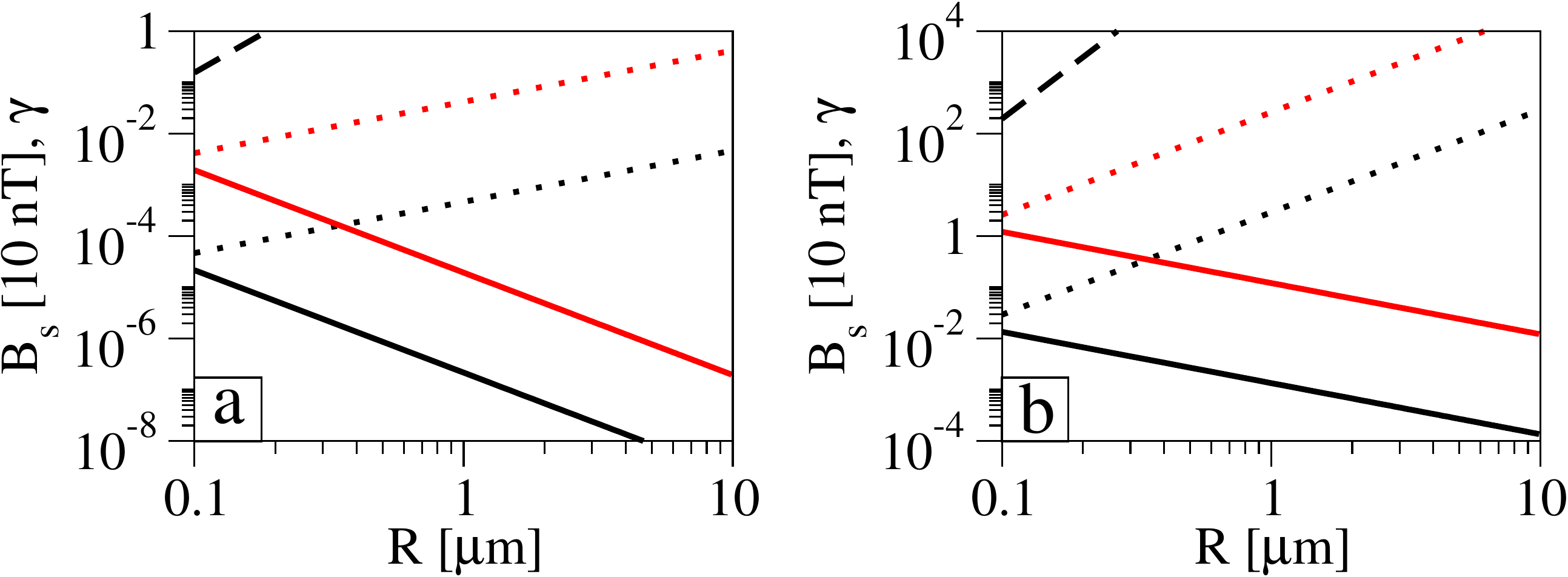}
\caption{(Color online) Figures of merit for an exchange based source (per one Mn ion) for n-GaAs (black) and p-ZnTe (red) in a) 2D and b) 1D. Apart from $2w=8$ nm and parameters given in Fig.~\ref{fig:dipole-compar}, we used $g_{\rm Mn}=2$, $I_{\rm Mn}=5/2$, $\beta=9$ meV nm$^3$ for n-GaAs, and $\beta=-15$ meV nm$^3$, and $g=2$ for p-ZnTe.\cite{venghaus1980:PRB}}
\label{fig:exchange-compar}
\end{figure}

Using a contact source, Eq.~\eqref{eq:perturbation exchange}, instead of a dipole source,  Eq.~\eqref{eq:perturbation dipolar}, amounts to replacing the source field flux, commented below Eq.~\eqref{eq:SNR1}, by $\Phi_s \sim \beta I /\mu_B g A_{3-d}(w)$ where $w$ is the transverse dimension of the medium (the half of the width of the heterostructure for 2DEG, the wire radius for a one-dimensional case). We get analogs of Eqs.~\eqref{eq:Bp1}-\eqref{eq:SNR1} in the form
\be
B_t \sim -\frac{\Omega_d }{4\pi \Omega_{3-d}}g \beta I \mu_B \mu_0 (w \lambda_p)^{d-3} \chi(R) \Lambda_d(2k_F\lambda_p),
\ee
and, with $g_I$ the source atom $g$-factor,
\be 
\gamma \sim \frac{\Omega_d}{\Omega_{3-d}} \frac{g}{g_I} \beta  (w \lambda_p)^{d-3} R^3 \chi(R) \Lambda_d(2k_F\lambda_p).
\ee
Using the result for $\chi$ in the non-interacting case we get
\be
B_t \sim \frac{\pi^{-1-d}\Omega_d}{16 d\Omega_{3-d}} g \beta I \frac{m}{\hbar^2} \mu_B \mu_0 \frac{(w \lambda_p)^{d-3}}{k_F^{2-d} R^{d}} \Lambda_d(2k_F\lambda_p),
\ee
where we again suppressed the oscillating factor, and 
\be 
\gamma \sim -\frac{1}{4\pi^d d} \frac{\Omega_d}{\Omega_{3-d}} \frac{g}{g_I} \beta I \frac{m}{\hbar^2} \frac{(w \lambda_p)^{d-3}}{R^{d-3} k_F^{2-d}} \Lambda_d(2k_F\lambda_p).
\ee
The illustrative values for two different semiconductor materials in one and two dimensions are shown in Fig.~\ref{fig:exchange-compar}. The signal for an exchange based source with a single Mn atom is comparable to a dipolar source considered before. However, in principle there might be many magnetic atoms on the source tip. On the other hand, an atomistically localized source requires  free access to the medium surface and is therefore not directly available for standard 2DEGs. The question then arises, how to enable this technique for a 2DEG. One possibility is to look for exposed (surface) two-dimensional gases, such as Shockley-Tamm states on metal surfaces\cite{crommie1993:S} or topological insulators.\cite{Hasan2010} On the other hand, cleaved edge samples\cite{cleaved_edge_Yacoby} allow one to access even the standard heterostructure 2DEG, such as the one in GaAs, which are of our primary concern. As the second note, we remind that our previous treatment 
of a static 
response assumes the source field is fixed, irrespective of the medium back-action. This requires that the atom moment itself is fixed.\cite{note_1}
 This can be achieved by a ferromagnetic or an antiferromagnetic coating of the tip.

\subsection{Refocusing }

\begin{figure}
\includegraphics[width=0.45\textwidth]{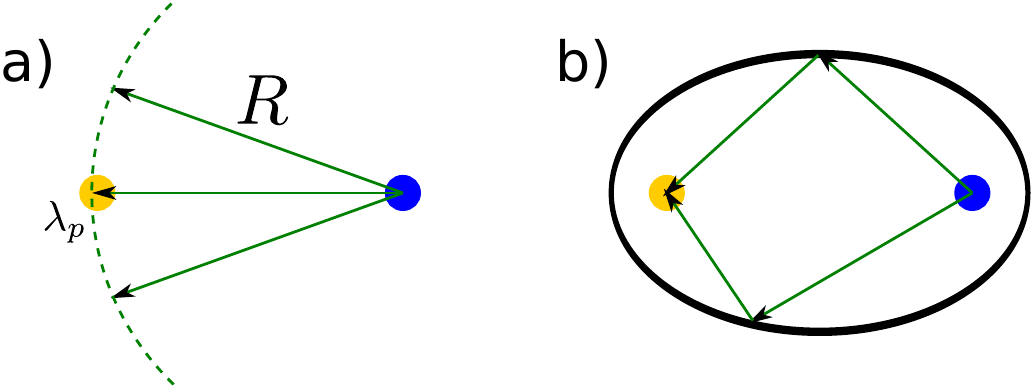}
\caption{(Color online) Illustration of a signal loss and its recovery by refocusing in two dimensions. a) In an open system, the probe (yellow) collects only a small fraction, of order $\lambda_p/R$, of the signal. b) With the source and the probe in the foci of an ellipse, all paths with a single reflection from a boundary have the same length, leading to a signal refocusing.}
\label{fig:refocusing}
\end{figure}

A different view on the origin of Eq.~\eqref{eq:non-int-ratio} is presented on the schematics in Fig.~\ref{fig:refocusing}a. Looking at the signal as emanating from the source, one half of it is collected at the probe in one dimension, while only a fraction of $\lambda_p/R$ in two dimensions, where most of the signal is lost. This leads us to consider possible geometries in which the lost signal could be recovered. Immediate examples are a parabolic antenna-receiver setup or source and probe in foci of an ellipse (see Fig.~\ref{fig:refocusing}b). That electrons in 2DEGs can be waveguided by top gates has been demonstrated.\cite{zumbuhl2002:PRL,frolov2009:N} Even though the time dependent picture we gave above is not completely adequate for a static response, and the susceptibility for shape-designed structures would need to be calculated, qualitatively we expect that the suppression given in Eq.~\eqref{eq:non-int-ratio}  can be removed by refocusing.  Taking again the distance of 1 micron, we estimate the 
refocusing would enhance the signal by an order of magnitude. 

\subsection{Interactions}

\label{sec:interactions}

\begin{figure}
\includegraphics[width=0.45\textwidth]{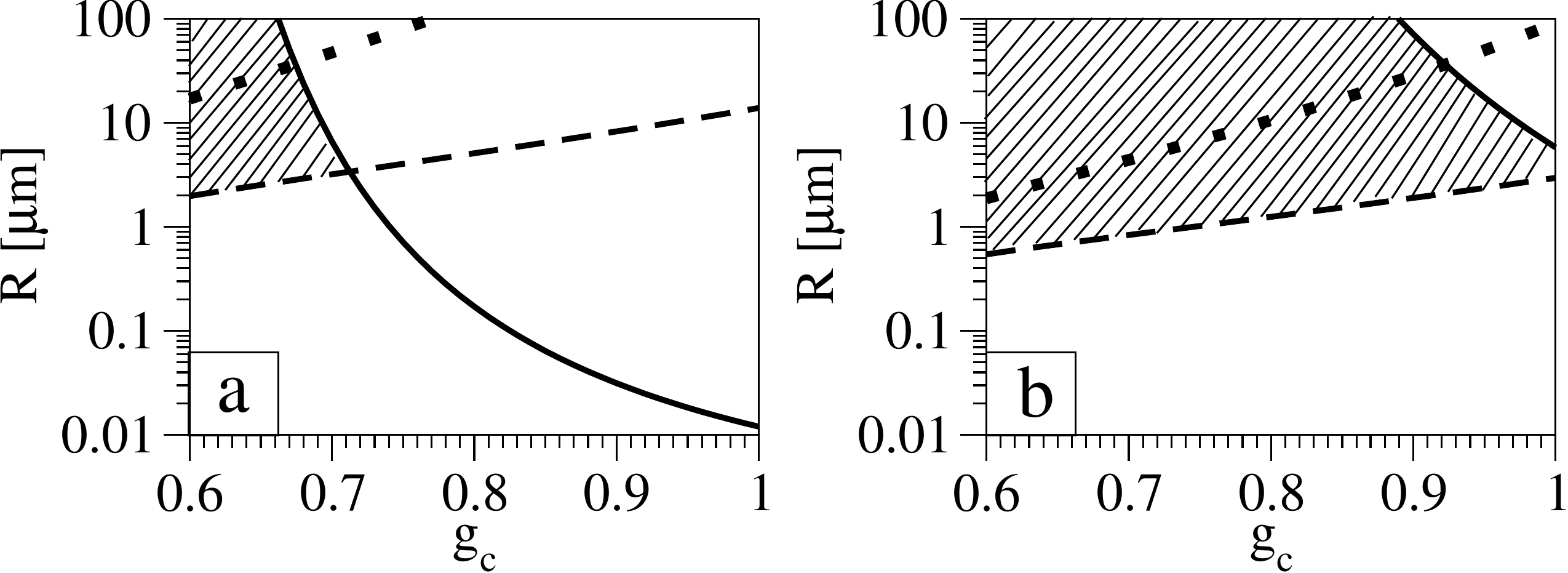}
\caption{The experimentally accessible region (shaded) in the area of the interaction strength (Luttinger parameter $g_c$, x axis) and source-probe distance ($R$, y axis). The region is defined by the area below the solid line ($B_t=10$ nT), and above the dashed ($\gamma=1$) or dotted ($\gamma^\prime=1$) line for a) GaAs and b) SiGe one-dimensional wire. We used $\lambda_p=10$ nm, $1/k_F=50$ nm. Further, $g=5.6$, $m=0.2m_e$ (SiGe) and $g=-0.44$, $m=0.067m_e$ (GaAs). With the density modulation lock-in technique the dashed line is irrelevant and the shaded area extends all the way below the solid line.
}
\label{fig:oneD-enhancement}
\end{figure}

Above we have discussed non-interacting systems.
Generally, the susceptibility is expected to be substantially enhanced by electron-electron interactions  in one\cite{egger1996:PRB,braunecker2009:PRB} and two dimensions.
\cite{chubukov2003:PRB,chubukov2005:PRB,simon2007:PRL,simon2008:PRB,chesi2009:PRB}
How exactly the influence looks like is one of the main motivations for studying the susceptibility experimentally. The uniform enhancement of the spin susceptibility can be understood as the renormalization of the effective mass and g-factor, with a good correspondence of the theory and experiments.\cite{depalo2005:PRL,zak2012b:PRB} The enhancement grows upon lowering the density, and close to the metal-insulator transition almost an order of magnitude enhancement has been seen in 2DEGs.\cite{zhu2003:PRL,vakili2004:PRL} Since we consider the low density regime, one order of magnitude enhancement for the signal magnitudes compared to what we plot for two-dimensional structures is reasonable. However, we expect that the local susceptibility might be boosted even stronger, e.g., at long distances. This is exemplified by a one-dimensional wire in the Luttinger liquid regime, for 
which the spin susceptibility has been derived as\cite{egger1996:PRB,braunecker2009:PRB}
\be
\chi(R) \approx -\frac{1}{4\pi a v_F \hbar} \frac{\Gamma(g_c-1/2)}{\Gamma(g_c)\Gamma(1/2)}\cos(2k_F R) (a/R)^{2g_c-1}.
\label{eq:luttinger}
\ee
Here $a$ is the lattice constant, $v_F$ the Fermi velocity, $\Gamma$ the Euler Gamma function, and $g_c$ the Luttinger interaction parameter for the charge sector. The non-interacting result, Eq.~\eqref{eq:non-interacting result}, is recovered for $g_c=1$. Smaller values of $g_c$ reflect stronger repulsive interactions, resulting in a slower decay of the susceptibility with the distance.
Using Eq.~\eqref{eq:luttinger} we plot in Fig.~\ref{fig:oneD-enhancement} the range for the source-probe distance where the susceptibility is experimentally easily accessible, as a function of $g_c$. Strong enhancement of the susceptibility with the interaction strength is apparent. The interaction can substantially improve the figures of merit for the susceptibility measurement, enhancing them by orders of magnitude.

\section{Carbon based materials}


\begin{figure}
\includegraphics[width=0.45\textwidth]{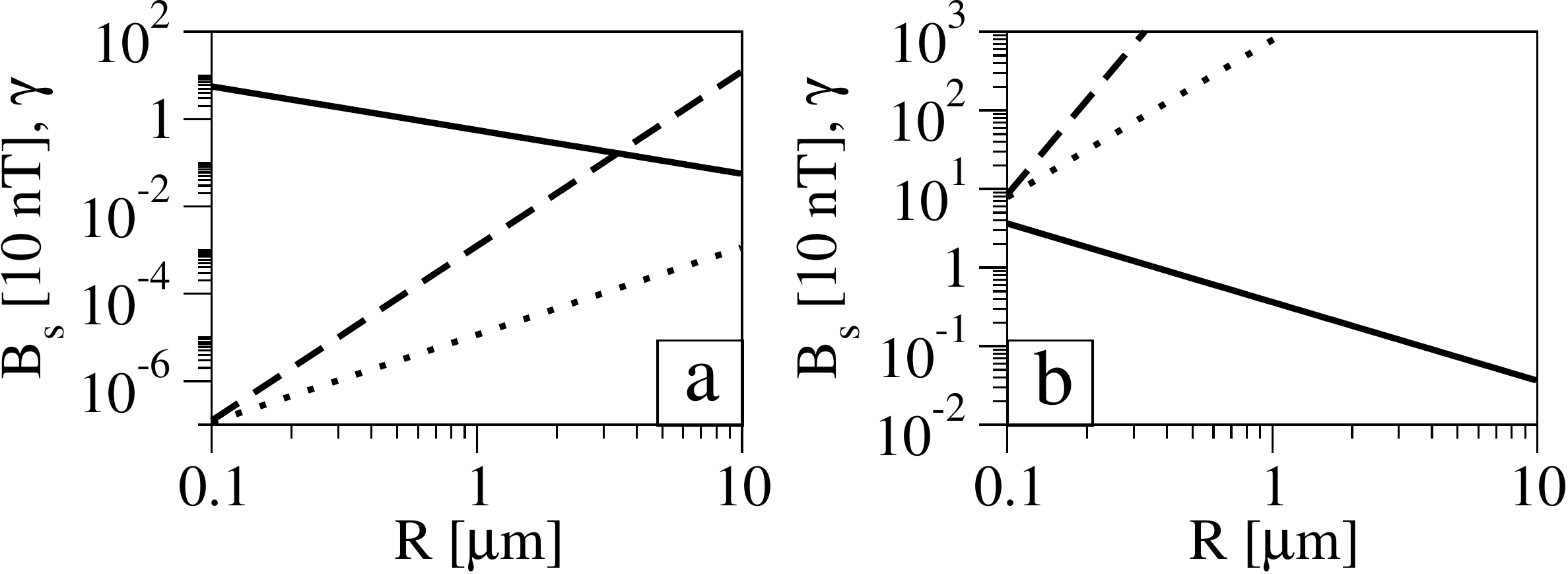}
\caption{Figures of merit for semiconducting carbon nanotube, $B_t/10$ nT (solid), $\gamma$ (dotted), and $\gamma^\prime$ (dashed), for a) dipolar b) exchange source per a single Mn atom. We used $\beta/v_0=4$ eV,\cite{dietl2001:PRB} $v_F=10^6$ m/s, 
$w=1$ nm, $k_F=1/50$ nm, $\lambda_p=5$ nm, and $g=2$.}
\label{fig:carbon}
\end{figure}

The carbon-based low-dimensional structures acquired a great deal of attention recently.\cite{dresselhaus_book, Novoselov_2005, Novoselov_2009} Apart from their present popularity, stemming from suggestions for their use as a basis for spintronics based devices as well as for spin qubits,\cite{cnt_ext_basel_2, Bulaev_2008, cnt_ext_delft, cnt_ext_kop, cnt_ext_kuemmeth,Churchill_2009,cnt_ext_basel_1,Guido_Nature, cnt_ext_delft_2,cnt_helical_2011,cnt_MF_2012, bilayer_MF_2011, bilayer_Marcus,bilayer_Yacoby,bilayer_Yacoby_2,ribbon_MF_2013} we are motivated by their predicted strong electron-electron interactions and potentially strong spin susceptibility signal magnitude.\cite{brey2007:PRL,kogan2011:PRB,kogan2012:U,RKKY_graphene_review,klinovaja2012:U} Namely, since the surface is exposed even in their two-dimensional form, they can be closely approached and are amenable also to the exchange-based (contact) sources. On the other hand, the helical character of graphene wavefunctions often leads to the 
susceptibility sign inversion upon moving between sublattices. For example, for metallic nanotubes $\chi^{AB}=-\chi^{AA}$,
where $A$ and $B$ denote the sublattices.\cite{klinovaja2012:U}  Since the probe we consider does not have an atomic resolution, the signal 
averages out to zero, and the susceptibility cannot be measured in this case.

The spin susceptibility for a single layer graphene at the Dirac point is given by\cite{kogan2011:PRB,brey2007:PRL}
\be
\chi(R) \approx \frac{1}{512 \hbar v_F R^3},
\label{eq:SL}
\ee
with $v_F$ the Fermi velocity. In a bilayer graphene\cite{kogan2011:PRB}
\be
\chi(R) \approx \frac{m}{32 \pi^2 \hbar^2 R^2},
\label{eq:BL}
\ee
with $m$ the effective mass (Ref.~\onlinecite{mayorov2010:S} found $m=0.029\,m_e$).  
In these formulas we averaged the susceptibility over the unit cell, so that fast oscillating terms, as well as the sublattice dependence, disappear. Doping a single layer graphene increases the susceptibility, changing its long distance fall-off from $1/R^3$ to $1/R^2$. However, this slower decaying contribution has the opposite sign on different sublattices, so that for our purposes there is little difference between doped and pristine graphene.\cite{note_4}
The expected signal in graphene (figure not shown) is well below the considered detection threshold.

We now consider a semiconducting carbon nanotube, where the susceptibility atom to atom sign oscillations are not present, and $\chi^{AA}=\chi^{AB}=\chi$, with\cite{klinovaja2012:U}
\be
\chi(R) = \frac{k_G}{4\pi \hbar v_F}\frac{\cos(2 k_F R)}{k_F R},
\label{eq:CNT}
\ee 
where $k_G=1/3 w$ and $w$ is the nanotube radius. We illustrate the expected signal in Fig.~\ref{fig:carbon}. Based on these numbers, we expect the signal to be measurable for both dipolar and exchange based probes, with further enhancement by interactions.

\section{Alternative setups}

\subsection{Pump glued to probe}


\begin{figure}
\includegraphics[width=0.45\textwidth]{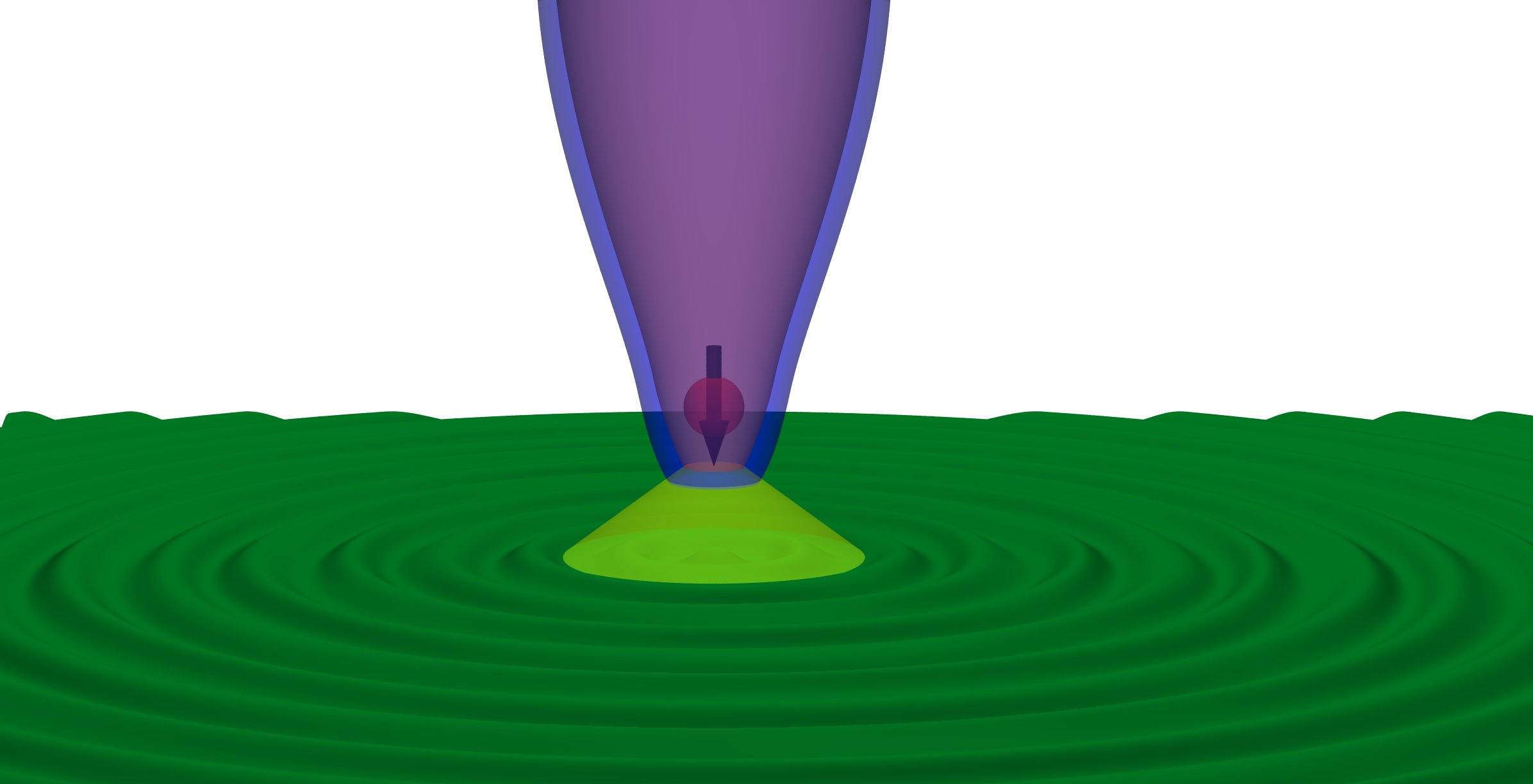}
\caption{(Color online) Single pillar hosts both the probe and the source, the latter as the (anti)ferromagnetic coating. The background field is not correlated with the probe distance from the medium. The short-distance structure of the susceptibility is accessible.}
\label{fig:one_probe_source} 
\end{figure}

\label{sec:glued}

\begin{figure}
\includegraphics[width=0.45\textwidth]{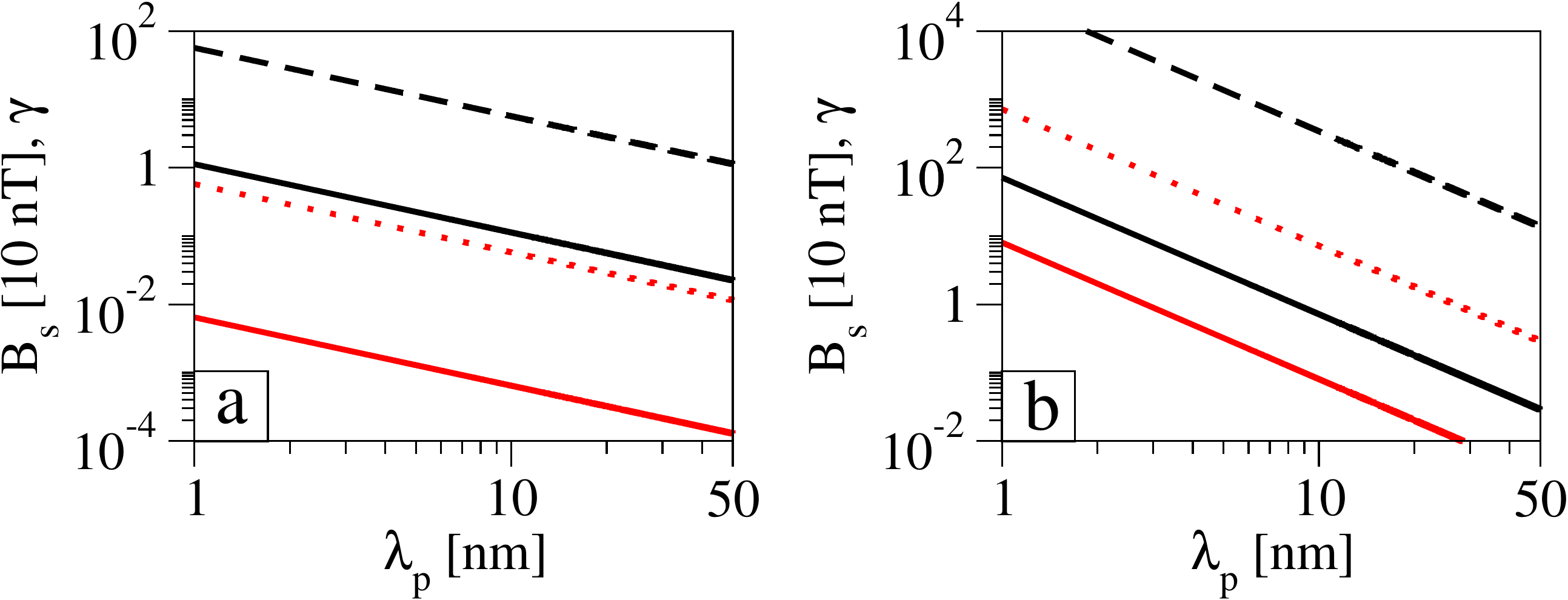}
\caption{(Color online) The signal for the single pillar setup in a) 2D and b) 1D geometry, for a dipolar [black; using Eq.~\eqref{eq:glued1}] and exchange [red; using Eq.~\eqref{eq:glued2}] source. Values plotted for GaAs (solid), InGaAs (dashed), p-ZnTe (dotted). We used parameters given in Figs.~\ref{fig:dipole-compar} and \ref{fig:exchange-compar} and set $\lambda_s=\lambda_p$.}
\label{fig:one_probe_source_data} 
\end{figure}

A possible way to suppress the background noise from a dipolar source is to consider the probe and source fixed on the same crystal. Their distance is then constant (up to negligible thermal mechanical vibrations), and so is the background field. 
We  evaluate the signal assuming that the distance between the probe or the source and the medium is smaller than the Fermi wavelength, $\lambda_p,\lambda_s \lesssim \pi/2k_F$,
and use the short distance limits for non-interacting particles susceptibility in one and two dimensions. For the dipole based source we get
\be
B_t \sim -B_r \frac{\Omega_d^2}{4\pi} \mu_0 g^2 \mu_B^2 \lambda_s^d \lambda_p^{d-3} \overline{\chi}(0),
\label{eq:glued1}
\ee
and for an exchange based source
\be
B_t \sim \frac{\Omega_d}{4\pi\Omega_{3-d}} \mu_0 g\mu_B \beta I (\lambda_p w)^{d-3} \overline{\chi}(0).
\label{eq:glued2}
\ee
In this formulas, the overline denotes averaging\cite{note_5}
\be
\overline{\chi}(R)=A^{-1} \int_{{\bf r}\in A} {\rm d}{\bf r}\, \chi(|R{\bf \hat{x}}+{\bf r}|),
\label{eq:A average}
\ee 
over the area $A$ with linear dimension of the order of $\lambda_p$ for $d=1$, and $1/k_F$ for $d=2$. The reason for this difference is again the saturation of the dipolar field in two dimensions once $\lambda_p$ falls below $1/k_F$. Equations \eqref{eq:glued1}-\eqref{eq:glued2} show that the single tip design, depicted in Fig.~\ref{fig:one_probe_source}, allows to access the susceptibility on short lengthscales, complementing the separate source-probe setup considered previously. We present the estimated signal strength in Fig.~\ref{fig:one_probe_source_data}.
The stronger signal in this setup arises from the fact that the spin susceptibility has its maximum close to $R=0$.
Growing several pillars with an NV center detector at each, displaced from the source tip in various distances, all on a single crystal, is another possibility of a reduced fluctuating background measurement.

\subsection{Spin noise}
\label{sec:noise}

\begin{figure}
\includegraphics[width=0.45\textwidth]{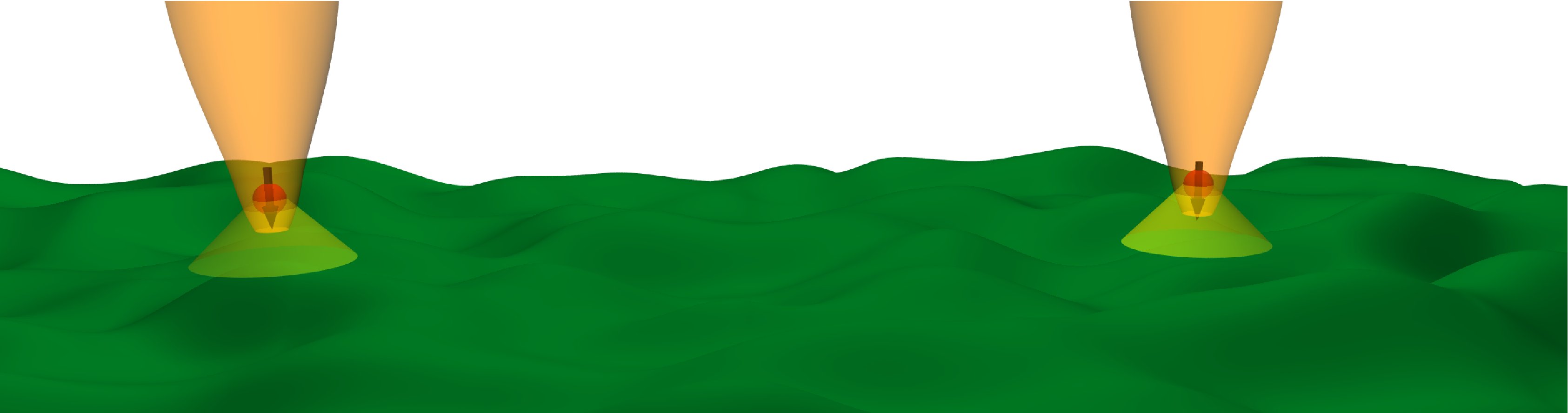}
\caption{(Color online) The spin susceptibility from noise. Equilibrium magnetization fluctuations (random waves) reflect the susceptibility through the fluctuation-dissipation theorem. Observing a single probe reveals $\chi({\bf r},{\bf r^\prime})$ for ${\bf r}\approx{\bf r^\prime}$, whereas noise cross-correlation of two distant probes is required to map the spin susceptibility in space coordinates.}
\label{fig:noise-fancy} 
\end{figure}

The spin susceptibility is related to the equilibrium noise via the fluctuation-dissipation theorem. This offers 
an interesting alternative to the source-probe setup.  Instead of the magnetization produced by the source, one can aim at the magnetization noise in the medium measured by a single probe or a pair of probes, as depicted schematically in Fig.~\ref{fig:noise-fancy}. We now estimate the signal in such a setup.

\begin{figure}
\includegraphics[width=0.45\textwidth]{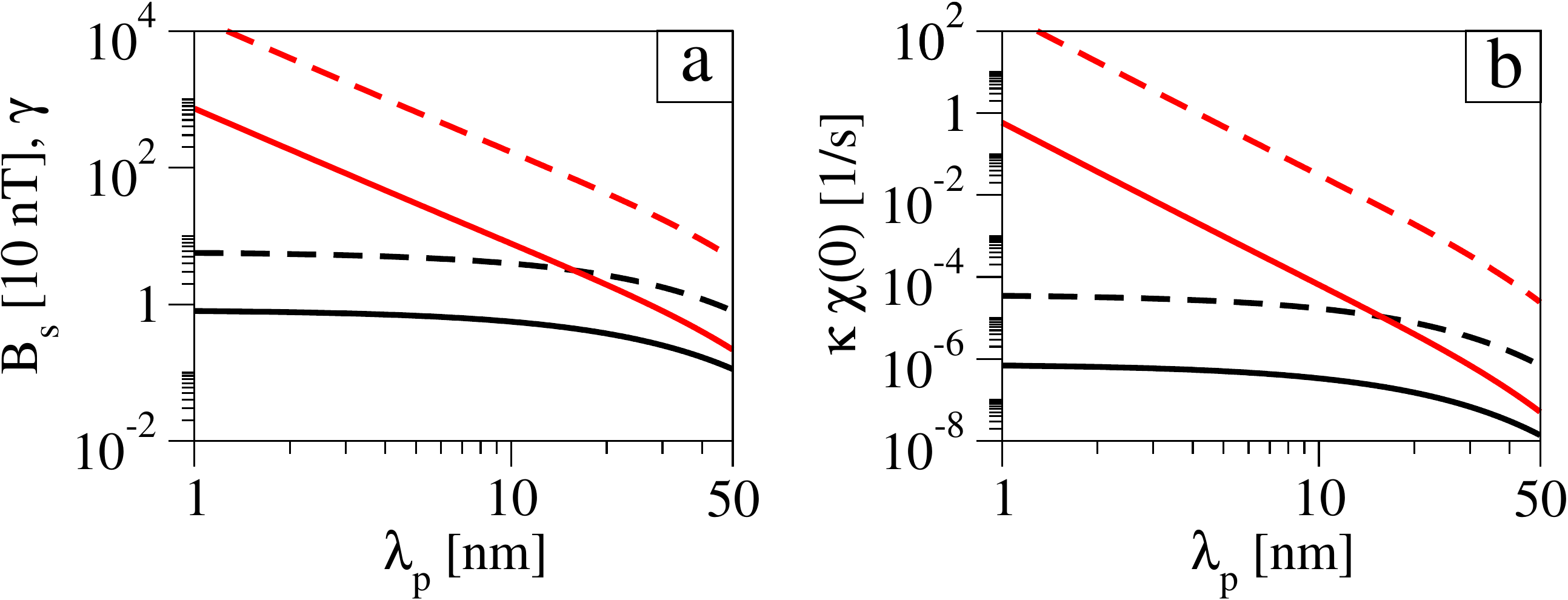}
\caption{(Color online) The expected level of noise in a 2D (black) and 1D (red) geometry in GaAs (solid) and InGaAs (dashed). a) Typical value of the noise induced magnetic field, Eq.~\eqref{eq:noise}. b) Phase correlator decay scale, $\kappa \overline{\chi}(0)$ using Eq.~\eqref{eq:correlation frequency}. Apart from parameters given in Fig.~\ref{fig:dipole-compar}, we used $\gamma=1$ meV, and $T=4$ K.}
\label{fig:noise data} 
\end{figure}

The fluctuation-dissipation theorem,
\be
2 \hbar \Im \chi_{\alpha \alpha}({\bf r},{\bf r^\prime};\omega) = [\exp(- \hbar \omega /k_B T)-1] S_{\alpha \alpha} ({\bf r},{\bf r^\prime};\omega),
\label{eq:fluctuation-dissipation}
\ee
relates the imaginary part of the susceptibility to the dynamical structure factor 
\be
S_{\alpha \alpha^\prime}({\bf r},{\bf r^\prime}; t-t^\prime) = \langle  
\rho^s_\alpha ({\bf r}, t) \rho^s_{\alpha^\prime} ({\bf r^\prime}, t^\prime)  \rangle.
\label{eq:structure factor}
\ee
Here $T$ is the temperature and $k_B$ the Boltzmann constant, and the averaging on the right hand side is with the the equilibrium density matrix.
The correlator of the probe(s) magnetic fields is 
\be
S_B(|{\bf r}-{\bf r'}|,t-t')=\langle B({\bf r},t) B({\bf r'},t')\rangle,
\ee
Assuming this correlator decays monotonically over a timescale $1/\gamma$, with $\hbar\gamma$ of the order of the bandwidth, formally defined as
\be
\gamma^{-1} = S_B(R,0)^{-1} \int_0^\infty {\rm d t}\, S_B(R,t),
\ee
we use the fluctuation-dissipation theorem to relate the susceptibility to the equal times correlator,
\be
S_B(R,0) \sim \left(\frac{\mu_0}{8 \pi}  \mu_B g \Omega_d \lambda_p^{d-3} \Lambda_d(2k_F \lambda_p)\right)^2 \overline{\chi}(R) E^*.
\label{eq:noise}
\ee
Here $E^*=\pi\hbar \gamma / 2{\rm ln}(\hbar \gamma/2k_B T)$ at low temperature, $\hbar \gamma \gg k_B T$, and $E^*=k_B T$ at high temperature, $\hbar \gamma \ll k_B T$.\cite{note_6}
The overline on the susceptibility denotes an average over an area with linear dimension $\lambda_p$, which gives $\overline{\chi}(R) \approx \chi(R)$ for $R \gg k_F \gtrsim \lambda_p$ and $\overline{\chi}(0)$ is understood as described in the previous section, below Eq.~\eqref{eq:A average}.
A typical magnitude of the field produced by the magnetization noise is then $B\sim \surd S_B(0,0)$, and we plot it in Fig.~\eqref{fig:noise data}a.

The previously derived field magnitude is a representative instantaneous value. Though typically well above the detection limit, such a comparison, as done in Fig.~\ref{fig:noise data}a, implicitly assumes the measurement has a time resolution below the noise decay-in-time scale, being $1/\gamma$. If we relax this time resolution requirement, we can proceed in the following way. One conceivable measurement is the noise induced decay of the phase autocorrelation (a single-probe measurement)
\be
A(t)=\langle \exp[{i} \Phi(0)] \exp[-{i} \Phi(t)] \rangle,
\ee
where the accumulated phase is
\be
\Phi(t) = (\mu_p/\hbar) \int_0^t B(t') {\rm d} t',
\ee
with $\mu_p$ the probe magnetic moment. Assuming, for simplicity, that the probability distribution of the magnetization fluctuations is Gaussian, we get
\be
\begin{split}
A(t) &= \exp \left( - \frac{\mu_p^2}{2\hbar^2} \int_{-t}^t {\rm d}t'\, S_B(0,t') (t-|t'|) \right)\\
& \approx \exp \left[ -(\mu_p/\hbar)^2 S_B(0,0)t^2 {\rm min} \{2/(\gamma t),1\} \right]. 
\end{split}
\label{eq:autocorrelation}
\ee
Assuming low time resolution $t\gg 1/\gamma$ and the low temperature limit for $E^*$, we finally get 
\be
A(t) = \exp \big( - \kappa \, \overline{\chi}(0) \, t\big),
\label{eq:autocorrelation result}
\ee
with
\be
\kappa \sim \mu_p^2 \pi \hbar^{-1} \left(\frac{\mu_0}{8 \pi}  \mu_B g \Omega_d \lambda_p^{d-3} \Lambda_d(2k_F \lambda_p)\right)^2.
\label{eq:correlation frequency}
\ee
The frequency scale $\kappa \overline{\chi}(0)$ is plotted in Fig.~\ref{fig:noise data}b.

To infer the susceptibility spatial dependence, we consider a measurement of the cross-correlation of phases accumulated over an identical interval of time of length $t$ in two probes positioned at relative distance $R$,
\be
C(R,t)=\langle \exp[{i} \Phi_1(t)] \exp[-{i} \Phi_2(t)] \rangle.
\ee
We get expressions analogous to those above, namely $C(t)$ is given by Eq.~\eqref{eq:autocorrelation} upon replacement $S_B(0,t') \to S_B(0,t')-S_B(R,t')$ and 
\be
C(R,t) = \exp \big( - \kappa  [\overline{\chi}(0) - \chi(R)] t\big),
\label{eq:crosscorrelation result}
\ee
with $\kappa$ given by Eq.~\eqref{eq:correlation frequency}.

The advantage of the noise measurement is that there is no need for a source, and consequently no accompanying background field. The disadvantage is that unless a time resolved measurement with the resolution below $1/\gamma$ is available, the susceptibility is given by Eq.~\eqref{eq:noise} with a factor $E^*$ that depends on the temperature and on not very well known characteristics of the noise fluctuations, such as the correlation scale $\gamma$ and possibly with non-Gaussian statistics character (higher moments in the correlators).


\section{Conclusions}

We investigated the feasibility of a magnetic spatially resolved measurements of the spin susceptibility of low-dimensional structures. We suggested to use a nanoscale magnetic sensor based on an NV center implanted in a nanopillar attached to an AFM tip. We compared the expected signal magnitude with the experimentally demonstrated detection limit. We suggest a two-tip (single-tip) source-probe measurement to access the long-range (short-range) structure of the spin susceptibility. We quantified the effectivity of a dipolar and an exchange based magnetic source. We also analyzed an alternative setup in which the susceptibility is extracted from the correlations in the noise, for which no source is necessary. We find that in one-dimensional systems, such as semiconducting nanowires and carbon nanotubes, the susceptibility is typically well within the current detection limits. The two-dimensional electron gases with high g-factors are most probably brought above the detection limit by interactions. 
In graphene 
we expect the signal to be too weak for a measurement, while the spin susceptibility of GaAs/AlGaAs 2DEGs might be detectable if interactions turn out to be strong enough and/or signal refocusing techniques are employed.

\acknowledgments

This work is supported by the Swiss NSF, NCCR Nanoscience, NCCR QSIT, and SCIEX. P.S. further acknowledges the support of COQI-APVV-0646-10. 

\appendix

\section{$\Lambda_d$ functions}

\begin{figure}
\includegraphics[width=0.45\textwidth]{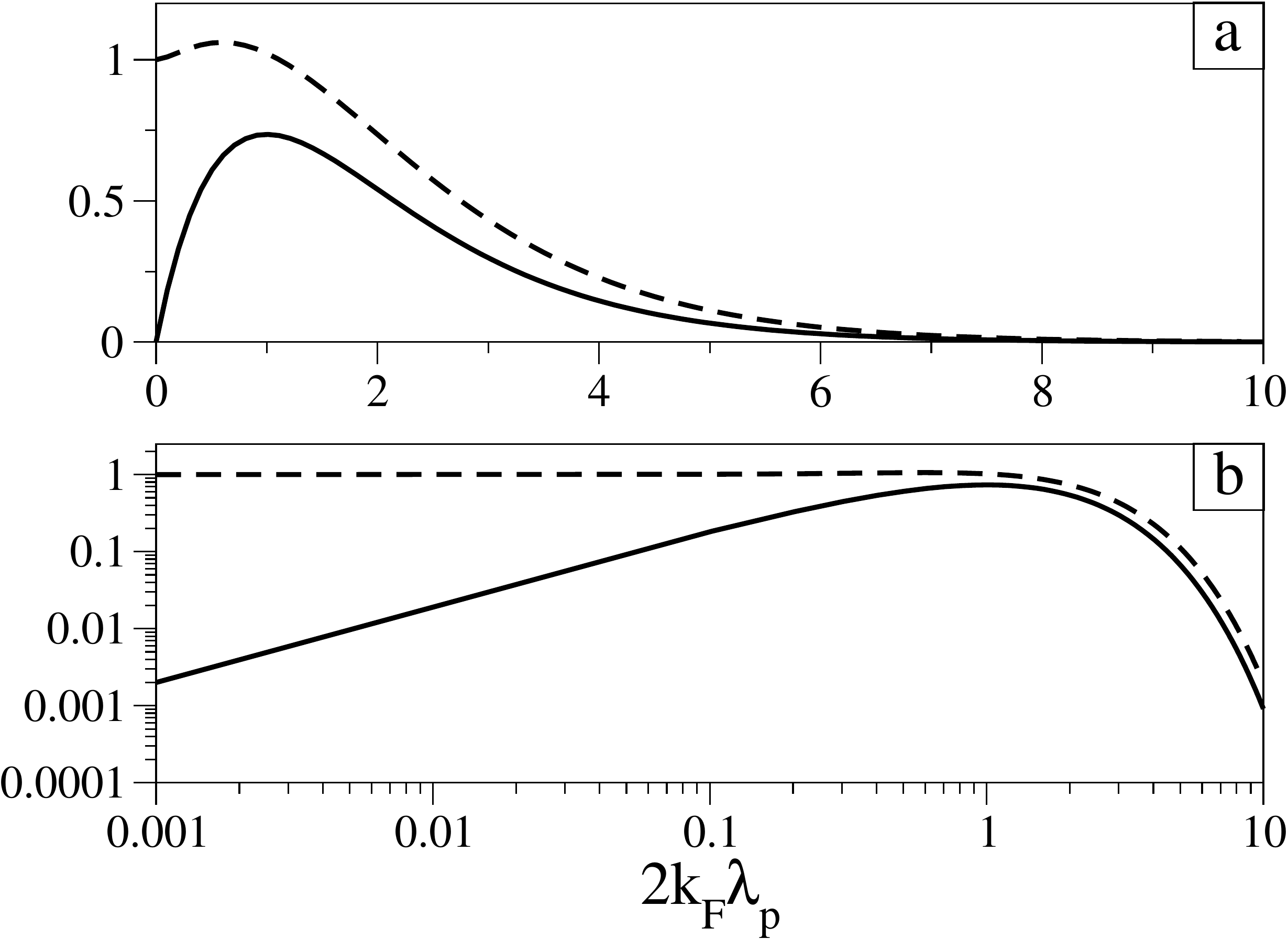}
\caption{The modulation functions $\Lambda_1$ (dashed) and $\Lambda_2$ (solid), which quantify the probe signal collection efficiency. The $x$ axis is in a) linear and b) logarithmic scale. }
\label{fig:Od functions} 
\end{figure}

\label{app:O}

To calculate the modulation functions $\Lambda_d$, we assume a source flux is localized below the spin susceptibility wavelength around the coordinate system origin and induces a magnetization profile $m(x,y) \approx m \cos[2 k_F (x-R)]$, which is a good approximation for $2k_F R \gg 1$. We define $\Lambda_d$ writing
\be
B_t ([R,0,\lambda_p]) = \frac{\mu_0}{4\pi} m \cos(2k_F R) \Omega_d \lambda_p^{d-3} \Lambda_d(2k_F \lambda_p).
\ee
The left hand side is given by the integral
\be
B_t({\bf r}) = \int {\rm d}{\bf r}^\prime T_{zz}({\bf r}-{\bf r}^\prime) m({\bf r}^\prime),
\ee
which can be analytically calculated with the assumed simplified form of the magnetization profile. We get $\Lambda_2(x)=2x\exp(x)$ and 
\be
\Lambda_1 (x) =\frac{2}{3}x^2 K_2(x) - \frac{2}{3}G^{2,1}_{1,3}\left(\frac{x^2}{4} 
\genfrac{|}{.}{0pt}{}{-1/2}{0,1,1/2} \right),
\ee
with $K_n$ the modified Bessel function of the second kind and $G$ the Meijer G-function. The small and large argument limits of $\Lambda_1$ are given in the main text and both
$\Lambda_1$ and $\Lambda_2$ 
are plotted in Fig.~\ref{fig:Od functions} for illustration.


\begin{thebibliography}{99}
\bibitem{note_2}This 
is 
closely related to the fact that an integral of a dipole field through any plane not containing the source dipole is zero, as the magnetic field lines close upon themselves. This would lead to an analogous suppression factor $\Lambda_d(2k_F\lambda_s)$ for a dipolar source. We solved this issue by considering a bar-like source magnet.
\bibitem{note_1} {Without this 
fixing the Hamiltonian in Eq.~\eqref{eq:perturbation exchange} preserves the total spin of the composite system of the atom and the medium, so that at maximum $2\hbar I$ of spin can be transferred to the medium.}
\bibitem{note_5}{The averaging is necessary as an integrable singularity of the susceptibility might be expected at $R=0$. This is the case for non-interacting particles in two dimensions, where $\chi(R) \sim \ln (k_F R)$ for $k_F R\ll 1$.}
\bibitem{note_6}{We used an exponential form of the correlator decay, to get explicit formulas. The results will remain qualitatively the same unless the correlator has anomalous behavior (oscillations or power law tails).}
\end{thebibliography}
\end{document}